# Compressed domain vibration detection and classification for distributed acoustic sensing

XINGLIANG SHEN,[1,2] HUAN WU,[3,*] KUN ZHU,[1] YUJIA LI,[4] HUA ZHENG,[1] JIALONG LI,[2] LIYANG SHAO,[2,*] PERRY PING SHUM,[2] CHAO LU[1]

[1] *Department of Electronic and Information Engineering, The Hong Kong Polytechnic University, Hong Kong, China*
[2] *Department of Electronic and Electrical Engineering, Southern University of Science and Technology, Shenzhen 518055, Guangdong, China*
[3] *Department of Land Surveying and Geo-Informatics, The Hong Kong Polytechnic University, Hong Kong, China*
[4] *Department of Electrical Engineering, The Hong Kong Polytechnic University, Hong Kong, China*
*Corresponding author: hkpolyu.wu@polyu.edu.hk; shaoly@sustech.edu.cn*

**Distributed acoustic sensing (DAS) is a novel enabling technology that can turn existing fibre optic networks to distributed acoustic sensors. However, it faces the challenges of transmitting, storing, and processing massive streams of data which are orders of magnitude larger than that collected from point sensors. The gap between intensive data generated by DAS and modern computing system with limited reading/writing speed and storage capacity imposes restrictions on many applications. Compressive sensing (CS) is a revolutionary signal acquisition method that allows a signal to be acquired and reconstructed with significantly fewer samples than that required by Nyquist-Shannon theorem. Though the data size is greatly reduced in the sampling stage, the reconstruction of the compressed data is however time and computation consuming. To address this challenge, we propose to map the feature extractor from Nyquist-domain to compressed-domain and therefore vibration detection and classification can be directly implemented in compressed-domain. The measured results show that our framework can be used to reduce the transmitted data size by 70% while achieves 99.4% true positive rate (TPR) and 0.04% false positive rate (TPR) along 5 km sensing fibre and 95.05% classification accuracy on a 5-class classification task.**

The distributed long-distance sensing mechanism provided by distributed acoustic sensing (DAS) opens new opportunities for many applications, such as seismic detection [1], pipeline surveillance [2], perimeter intrusion detection [3], submarine monitoring [4], etc. However, the densely distributed long-distance sensing capability coupled with high frequency sampling rate leads to huge flow of data, which puts a large burden on input/output interface (I/O), storage, and computation. For example, in seismic event detection, 128 TB of raw data is generated in 3 months when using 27 km dark fibre with 12000 channels and 500 Hz sampling rate for measurement [1].

Traditional analog-to-digital converting (ADC) is based on the Nyquist-Shannon sampling theorem, which requires the sampling rate to be at least twice the maximum frequency component available in the signal. However, this method heavily over sample the signal and leads to large storage and bandwidth requirements. To overcome this problem, a well-known sampling theorem called compressive sensing (CS) has been proposed in [5]. Assuming a real-valued, finite-length, and discrete-time signal $x$, which can be expressed as an $N \times 1$ column vector in $\mathbb{R}^N$. Any signal in $\mathbb{R}^N$ can be changed to a new basis as follows,

$$x = \sum_{i=1}^{N} s_i \boldsymbol{\psi}_i = \boldsymbol{\psi} s \quad (1)$$

where $\boldsymbol{\psi}$ is an $N \times N$ orthonormal basis. $s$ is $N \times 1$ vector of weighting coefficients. Obviously, $x$ and $s$ are equivalent representations of the signal in time domain and $\boldsymbol{\psi}$ domain, respectively. If the signal is a linear combination of only K non-zero coefficients (K $\ll$ N), the signal is compressible in $\boldsymbol{\psi}$ domain. Most natural signals are sparse in transform domain, for example, sounds can be sparsely represented in frequency domain and images in wavelet domain [6]. In traditional compression methods, the whole $N \times 1$ signal is first acquired and then K largest coefficients are located and reserved while the remaining $(N-K)$ coefficients are discarded. To address the inefficiency of sample-then-compress framework, CS compresses the signal in the sampling stage. A fixed observation matrix $\boldsymbol{\phi}$, independent from $\boldsymbol{\psi}$ is introduced to transform $x$ to the compressed signal $y$:

$$y = \boldsymbol{\phi} x = \boldsymbol{\phi} \boldsymbol{\psi} s \quad (2)$$

where $\boldsymbol{\phi}$ is a $M \times N$ matrix (M $\ll$ N) and $y$ is a $M \times 1$ vector. It is obvious the K-sparse characteristic is reserved in $y$ after the dimension reduction from $x$. Thus, $x$ can be recovered from only $M \approx K$ measurements by reconstruction algorithms.

The capability of reducing the data size directly in the sampling stage makes CS attractive to DAS. The first work that uses CS to reduce the data size is proposed by S. Qu et al [7]. To break the relationship between sensing distance and sampling rate trade-off in a simple DAS scheme, they also propose to use the CS sampling theory for data acquisition. An 8 kHz signal is reconstructed with average pulse repetition frequency less than 2 kHz [8]. This CS based sampling scheme is also demonstrated in multi-frequency DAS and OFDR based DAS [9,10]. Though CS greatly reduces the data size in the sampling stage, the incomplete set of measurements $y$ is reconstructed with complex reconstruction algorithm Orthogonal Matching Pursuit (OMP) for further processing in the above-mentioned works [7-10].

Fig. 1(a) (blue line) shows a vibration signal collected from our DAS system with 10 kHz pulse repetition rate [11]. After transforming to the Fourier domain by discrete Fourier transform (DFT), the normalized and sorted transformation coefficients are shown in Fig. 1(b). The 2143rd coefficient contains only 1% information of the maximum coefficient, which indicates the signal sparsity in DFT domain. Fig. 1(c) shows the compressed signal measured with an observation matrix with 30% measurement ratio (MR, defined as M/N). The reconstructed signal by OMP algorithm from Fig. 1(c) with K = 2143 are shown in Fig. 1(a) in green line. The Pearson Correlation Coefficient (PCC) of the original signal and reconstructed signal is 0.9086. The reconstruction performance of OMP algorithm under different MR and K regarding similarity and recovering time is shown in Fig. 2. When the MR is lower than 30%, the compression is very lossy. The PCC values are also proportional to sparsity coefficient K. The result shows that K should be as large as 2143 with at least 30% MR to achieve over 0.9 PCC. However, such large K and MR will greatly increase the reconstruction time. It is because there is no unique solution to the reconstructed signal when the number of measurements M is much larger than its original dimension N. Therefore, iterative optimization such as OMP is required. Fig 2 (b) shows the reconstruction time with different MR and K. We use MATLAB to implement the OMP on a computer with i7-10700 CPU and 16 GB RAM. With 30% MR and K = 24, the reconstruction time is 3.47 seconds. Since the computational complexity of the OMP algorithm is linear to the K [12], more than 3 minutes are needed to reconstruct an 8000 measurement points signal with K = 2143 and 30% MR.

Since signal reconstruction for compressed data is time consuming and computationally intensive, in this Letter, we propose a two-stage algorithm to locate and classify the DAS signal directly in the compressed-domain. In DAS, most positions are not vibrated, therefore, vibration detection is first conducted along the fibre with data collected by CS sampling method in Stage-I. Then, the detected vibration signals are processed by compressed-domain pattern recognition algorithm to classify the vibration event. Frequency band energy (FBE) that reflects both intensity and frequency information of a signal is selected as feature vector in both stages. For the DAS signal acquired by Nyquist-Shannon sampling method, FBEs are equal to the summation of filtered signal in the frequency bands, as expressed in Equation (3). Frequency-domain multiplication is equivalent to time-domain convolution. To develop the compressed-domain analysis, convolution is converted to matrix multiplication as expressed in Equation (4).

$$FBE = sum(|H|) = sum(|F \cdot X|) \quad (3)$$

$$H = F \cdot X \Leftrightarrow h = f \otimes x \Leftrightarrow h = Ax \quad (4)$$

where $h$, $f$ and $x$ are the filtered signal, Finite Impulse Response (FIR) bandpass filter (BPF) and input signal in time domain while $H$, $F$, $X$ are their Laplace transform. All of them are N × 1 vectors. Circulant matrix $A$ is an N × N Toeplitz matrix based on $f$, in which all row vectors are composed of the same elements and each row vector is rotated one element to the right relative to the preceding row vector. To construct the compressed-domain BPF (C-BPF), we can project the circulant matrix $A$ with observation matrix $\phi$. We aim to find a matrix $A_c$ that gives the following relationship [13]:

$$\phi A x = A_c \phi x \Rightarrow \phi A = A_c \phi \quad (6)$$

where $A_c$ is the M × M Toeplitz matrix, which corresponding to the C-BPF time domain. Because $\phi$ is not a square matrix, there is no true inverse of $\phi$, the right pseudo-inverse is used to solve the equation:

$$A_c = \phi A \phi^T (\phi \phi^T)^{-1} \quad (7)$$

Finally, we can get the compressed filter $f_c$ as the first row of $A_c$ and transfer it into frequency domain, $F_c$. The C-FBE can be expressed using $F_c$ and the Laplace transform of $y(Y)$ as:

$$C\text{-}FBE = sum(|F_c \cdot Y|) \quad (8)$$

To verify the proposed method, we collect a dataset with the same setup in [11]. The pulse width is 100 ns and pulse repetition frequency is 10 kHz. The vibration is applied through a speaker with 10 m fibre stick on it at around 5.02 km. Coherent detection is used to demodulate both intensity and phase information. The digital demodulation algorithm is implemented on a field programmable gate array (FPGA) chip (Zynq-7100, Xilinx Inc.). Four different events were created by playing soundtracks of thunderstorm (TH), welding (WD), jackhammer (JH), and shoveling (SH). The number of each signal clips is 40. Each signal clip lasts 3 seconds. To compare the vibration detection and classification performances of Nyquist-domain signal (Nyquist-Shannon sampled) and compressed-domain signal (CS sampled), we firstly use uniform pulse sequence to generate the original dataset and then multiply it with observation matrix to generate the compressed-domain dataset.

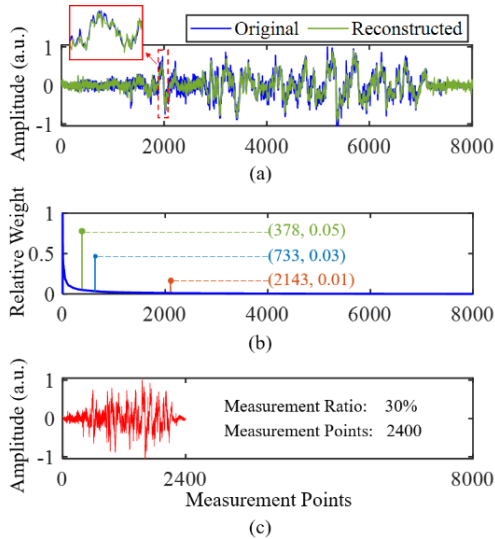

**Fig. 1.** (a)Original (blue) and reconstructed (green) signal, (b) sorted relative weight in DFT domain, and (c) compressed signal with 30% measurement ratio.

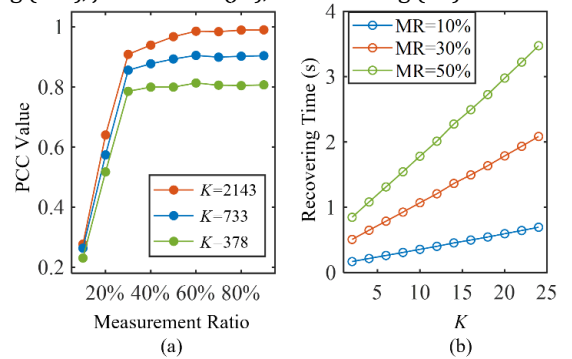

**Fig. 2.** (a) PCC value between original signal and reconstructed signal with different K and MR. (b) Recovering time of OMP algorithm with different K and MR.

The workflow of vibration detection and classification in Nyquist- and compressed-domain are shown in Fig. 3. For Nyquist-domain signals, 50 phase FBEs ($FBEs_P$) and 50 intensity FBEs ($FBEs_I$) are first extracted, respectively. According to the observation, most dominant frequencies are in low frequency bands, therefore, 50 BPFs cover 0-1500 Hz with 30 Hz bandwidth are constructed for FBE extraction. After feature extraction, vibration positions are detected based on threshold method. In Stage-I, only phase information is used. It is because the phase amplitude has a linear relationship with external vibration strength [14]. Signal detection based on short-time energy is a classic method proposed by Urkowitz in 1967[15]. The basic idea is that with the presence of a signal, the energy would be significantly larger compared with no signal present. $\lambda_{N1}$ represents the threshold for each FBE, if 80% of the FBEs surpass $\lambda_{N1}$, it is regarded as vibration signal. Otherwise, it is determined as non-vibration. The detected vibration signals are then classified in Stage-II. In this stage, both intensity and phase information are used as features since classification is based on the frequency characteristic of the signals. Phase and intensity FBEs are normalized and concatenated as $1 \times 100$ FBE vector as the input of support vector machine (SVM) classifier. For compressed-domain signals, the workflow is similar with that in Nyquist-domain except that the extracted feature is C-FBE constructed according to Equation (8).

Fig. 4 (a) shows a DAS signal along 5.2 km fibre with 0.8 second duration in Nyquist-domain. The vibration is applied at 5.02km. The corresponding compressed-domain signal with 30% MR is shown in Fig. 4(b). For better visualization, the signal is four times down-sampled along the fibre. It is obvious that vibration position (red line) has much higher amplitude than non-vibration positions (blue line). Fig. 4 (c) and (d) show the $FBEs_P$ and $C\text{-}FBEs_P$ along the sensing fibre. In both Nyquist- and compressed-domain, the energy level is an important feature that separate the vibration and non-vibration positions. The thresholds of $\lambda_{N1}$ and $\lambda_{C1}$ are set to different multiples of $FBEs_P$ and $C\text{-}FBEs_P$ without any vibration events in Nyquist- and compressed-domain, respectively. To evaluate the performance vibration detection, area under curve-receiver operating characteristic (AUC-ROC) curve is calculated, as shown in Fig. 5. ROC is a probability curve that describes the binary classification performance at various thresholds. The higher the AUC, the better the separation between vibration and non-vibration class. The AUC-ROC curve is plotted with true positive rate (TPR) against false positive rate (FAR), where FAR is on the x-axis and TPR is on the y-axis. The AUC of Nyquist-domain signal is 0.9806 and that of compressed-domain is 0.9807, which suggests that they have very close detection performance. Thresholds at the marked blue and red dots that with the minimum distance to the top left are chosen as system parameters in the following work. Specifically, the thresholds are 2.26 times of average $FBEs_P$ and 3.28 times of average $C\text{-}FBEs_P$ without external vibrations. In Nyquist-domain, the corresponding TPR is 98.8% and FAR is 0.07% while in compressed-domain, the corresponding TPR is 99.4% and FAR is 0.04%.

After discriminating the vibration positions along the fibre, true positive signals and false alarm signals are classified by multi-class classification support vector machine (SVM) in Stage-II. False alarm caused by environmental noise (EN) and detected 4 kinds of vibration signals are classed by a 5-class classification SVM models. We verified that classification accuracy can be greatly improved by stacking the phase/intensity information and expanding the training dataset by data augmentation [11]. Therefore, extracted phase and intensity FBEs and C-FBEs are both normalized and concatenated to 1×100 feature vectors for classification. Three data augmentation methods including time shifting, speed stretching, and pitch change are used to enlarge the training data size. The training dataset are expanded by 9 times by randomly generate the augmentation factors. The dataset is split into 5 folds and are evaluated with 5-fold cross validation to test their generalization capability. The confusion matrix yielded by the SVM classifier in Nyquist- and compressed- domain are provided in Fig. 6 (a) and (b). The y-axis shows the true label while the x-axis is the predicted label. Classification accuracy in each class is improved in compressed-domain. The overall classification accuracy in Nyquist-domain is 85.17% and that of compressed-domain can reach 95.05%. The higher classification accuracy in compressed-domain is mainly due to fewer EN samples from Stage-I. The 4-class event classification accuracy in Nyquist-domain is 97.74% and in compressed-domain is 97.78%, which indicates that compressed-domain signal has similar performance with that in Nyquist-domain. To visualize the feature vectors, T-Distributed Stochastic Neighbor Embedding (t-SNE) is used to embed the feature vectors into 2D plane as shown in Fig. 6 (c) and (d). The t-SNEs also verify that signal in Compressed-domain has similar feature vector representation capability.

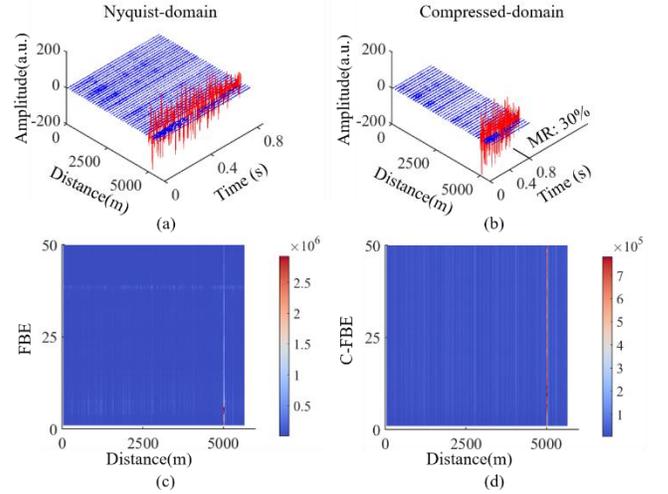

**Fig. 4.** DAS signal along 5.2 km fibre with 0.8 second duration with vibration applied at 5.02 km in (a) Nyquist-domain, (b) Compressed-domain with 30% MR. Corresponding (c) FBE and (d) C-FBE along the fibre. (Signal are four times down-sampled along the fibre for better visualization).

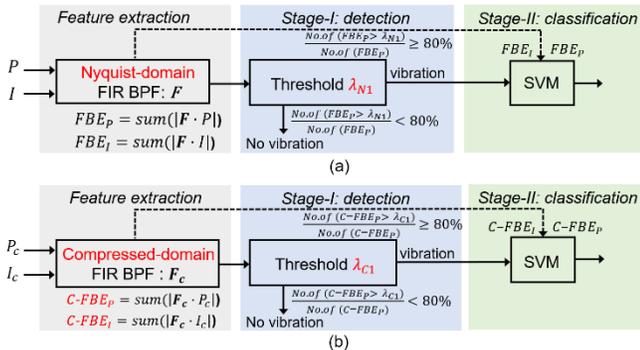

**Fig. 3.** Workflow of vibration detection and classification in (a) Nyquist-domain, (b) Compressed-domain.

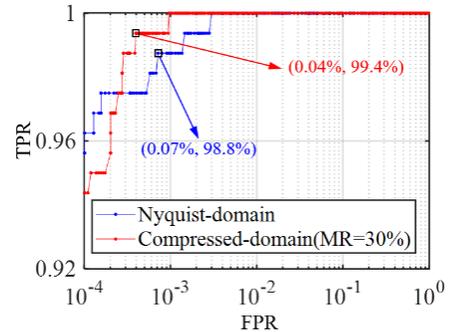

**Fig. 5.** AUC-ROC curves of Nyquist-domain (blue) and compressed-domain with 30% MR (red). The marked points are adopted thresholds for Stage-II.

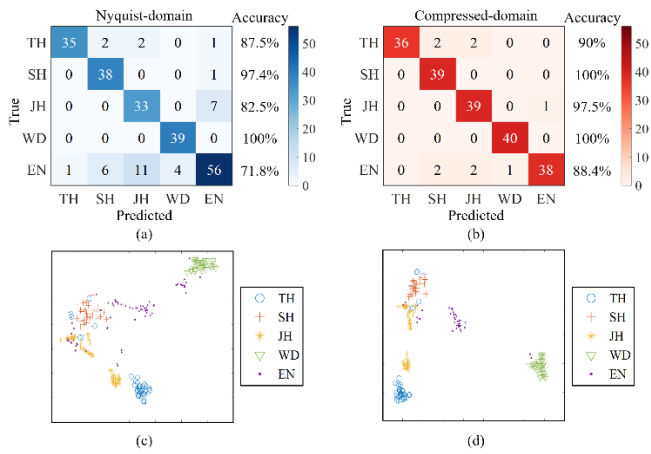

**Fig. 5.** Confusion matrices of the classification results in (a) Nyquist-domain, (b) compressed-domain and the respective embedded feature vectors in (c) Nyquist-domain, (d) compressed-domain.

In conclusion, efficient signal representation is the key to turning raw data into useful information. Instead of reconstructing the compressed signal and then extract the feature, we map the feature extractor from Nyquist-domain to compressed-domain for signal representation. In this Letter, for the first time, we report a compressed-domain vibration detection and classification method for DAS. A compressed-domain two-stage algorithm that uses frequency band energy as feature vector is demonstrated on a DAS system with 5.2km sensing fibre. The proposed method achieves 99.4% TPR and 0.04% FPR on vibration detection and 95.05% accuracy on a 5-class classification task with only 30% of data required by Nyquist-Shannon sampling.

**Funding.** Stable Support Program for Higher Education Institutions from Shenzhen Science, Technology & Innovation Commission (20200925162216001); Special Funds for the Major Fields of Colleges and Universities by the Department of Education of Guangdong Province (2021ZDZX1023); GuangDong Basic and Applied Basic Research Foundation (2021B1515120013); Open Fund of State Key Laboratory of Information Photonics and Optical Communications (Beijing University of Posts and Telecommunications), P. R. China (IPOC2020A002).

**Disclosures.** The authors declare no conflicts of interest.

**Data availability.** Data underlying the results presented in this paper will be published on the project website.